\documentclass[%
 aip,
 amsmath,amssymb,
 reprint,%
]{revtex4-1}

\usepackage{graphicx}
\usepackage{dcolumn}
\usepackage{bm}
\usepackage{orcidlink}
\usepackage[utf8]{inputenc}
\usepackage[T1]{fontenc}
\usepackage{mathptmx}
\usepackage{etoolbox}

\makeatletter
\def\@email#1#2{%
 \endgroup
 \patchcmd{\titleblock@produce}
  {\frontmatter@RRAPformat}
  {\frontmatter@RRAPformat{\produce@RRAP{*#1\href{mailto:#2}{#2}}}\frontmatter@RRAPformat}
  {}{}
}%
\makeatother
\begin{document}

\preprint{AIP/123-QED}

\title{High-precision measurement of the complex magneto-optical Kerr effect using weak measurement}
\author{Tong Li\orcidlink{0000-0001-8928-035X}}
\altaffiliation{These authors contributed equally to this work.}
 \affiliation{College of Physics, Sichuan University, Chengdu, Sichuan 610065, China}

\author{Yunhan Wang\orcidlink{0000-0001-8823-0326}}
 \altaffiliation{These authors contributed equally to this work.}
 \affiliation{College of Physics, Sichuan University, Chengdu, Sichuan 610065, China} 

\author{Yinghang Jiang\orcidlink{0000-0001-5312-2392}}
 \affiliation{College of Physics, Sichuan University, Chengdu, Sichuan 610065, China} 
 
\author{Sijie Zhang\orcidlink{0000-0002-6167-8339}}
 \affiliation{College of Physics, Sichuan University, Chengdu, Sichuan 610065, China}
 \affiliation{Guizhou University of Engineering Science, Bijie, Guizhou 551700, China}
 
\author{Lan Luo\orcidlink{0000-0002-6560-787X}}
  \email[Corresponding author: ]{naloul@163.com}
 \affiliation{the National Key Laboratory of Optical Field Manipulation
Science and Technology, Chinese Academy of Sciences, Chengdu 610209,
China}
 \affiliation{Key Laboratory of Science and Technology on Space Optoelectronic Precision Measurement, Chinese Academy of Sciences, Chengdu, Sichuan 610209, China}
 \affiliation{Institute of Optics and Electronics, Chinese Academy of Sciences, Chengdu, Sichuan  610209, China}
 
\author{Zhiyou Zhang\orcidlink{0000-0002-8619-9529}}
  \email[Corresponding author: ]{zhangzhiyou@scu.edu.cn}
 \affiliation{College of Physics, Sichuan University, Chengdu, Sichuan 610065, China}


\begin{abstract}
The present paper introduces a quantum weak measurement (WM) scheme for the measurement of the complex magneto-optical Kerr effect (MOKE). We achieve the simultaneous measurement of the Kerr rotation angle and the ellipticity in a single WM process by utilizing {two auxiliary pointers} derived from the same meter state. The experimental measurement precision for {both} the Kerr rotation angle and the ellipticity is capable of reaching $10^{-4}$ {deg}. This technique is also employed for the determination of the complex magneto-optical constant $Q$. The proposed method overcomes the limitation of acquiring the complex magneto-optical Kerr parameters through a multi-step measurement process, which was previously encountered. This breakthrough holds immense significance for efficiently measuring and applying the complex MOKE with high precision and cost-effectiveness.
\end{abstract}

\maketitle

The magneto-optical Kerr effect (MOKE), which originates from the anisotropy of the dielectric tensor\cite{Qiu:2014}, refers to the phenomenon that when a linear polarized beam incident on the interface of a magnetic medium, the polarization state or the intensity of the reflected light is changed. The corresponding magneto-optical Jones matrix used to characterize this effect can be represent as
\begin{equation}\label{eq2}
\Re_1=\left[\begin{array}{cc}
r_\mathrm{p p} & r_\mathrm{p s} \\
r_\mathrm{s p} & r_\mathrm{s s}
\end{array}\right],
\end{equation}
where $r_\mathrm{p p}$, $r_\mathrm{p s}$, $r_\mathrm{s p}$ and $r_\mathrm{s s}$ are the Fresnel coefficients of the magnetic medium \cite{You:1998}. There are three cases of MOKE depending on the orientation of the magnetization relative to the interface, the polar magneto-optical Kerr effect (PMOKE), the longitudinal magneto-optical Kerr effect (LMOKE), and the transverse magneto-optical Kerr effect (TMOKE) \cite{Yang:1993,You:1998}. Among them, both LMOKE and PMOKE cause changes in the polarization state of the reflected light. This process can be expressed by the complex magneto-optical Kerr angle 
\begin{equation}
\Theta_\mathrm{K}=\theta_\mathrm{K}-i\varepsilon_\mathrm{K},
\end{equation}
where the real and the imaginary part of it represent the Kerr optical rotation angle and the Kerr ellipticity of the reflected polarization, respectively. Without loss of generality, for horizontal polarization $\left|H \right\rangle$, it gives $\Theta_\mathrm{K}^\mathrm{H}={r_{\mathrm{sp} }}/{r_{\mathrm{pp} }} $, while for vertical polarization $\left|V \right\rangle$, it becomes $\Theta_\mathrm{K}^\mathrm{V}={r_{\mathrm{ps} }}/{r_{\mathrm{ss} }} $\cite{You:1998}. 

With the development of spintronics, MOKE especially the longitudinal and the polar configuration has shown the superiority for the in-situ investigation of the magnetic properties of magneto-optical devices and thin films, and has been widely applied in magnetism measurement \cite{Zak:1991}, magnetic domain imaging \cite{Williams:1951,Fowler:1952}, magneto-optical ellipsometry \cite{Berger:1997,Mok:2011}, magneto-optical recording \cite{Imamura:1985}, etc., because $\theta_\mathrm{K}$ and $\varepsilon_\mathrm{K}$ are highly sensitive to the magnetization $\bf M$ of the magnetic materials. 

In recent years, some studies used weak measurement (WM) to measure the {tiny} MOKE signals of {separate} $\theta_\mathrm{K}$ and $\varepsilon_\mathrm{K}$ \cite{He:21,Li:2020,Wang:2020,Luo:23,LiTong:23}, which has the advantages of {high precision and cost-effectiveness}. WM is a technique for the measurement of high-precision parameters by amplifying the observables. It includes the processes of pre-selection, weak coupling, and post-selection. It was first proposed by Aharonov, Albert, and Vaidman in 1988 \cite{Aharonov:1988}. Since WM improves the signal-to-noise ratio significantly and has extremely high sensitivity, it has been used in the detection of beam shift \cite{Hosten:2008,Zhou:2012,Ling:2023,Chen:2023}, frequency \cite{Starling:2010}, optical phases \cite{Liu:2022,Xu:2013,Luo:2022}, chirality \cite{Lidongmei:2018,Xiao:2021,wang23}, chemical reaction processes \cite{Ruisi:2020,Zhao:2022}, angular rotations \cite{Magana:2014}, refractive index \cite{Chen:2021,Zhou:2018}, optical conductivity \cite{Chen:2020}, and phase transition process \cite{Tang:2022}{, etc. }

To study the complex MOKE and apply it to the investigations of materials' properties, e.g., the measurement of the complex optical parameters, both $\theta_\mathrm{K}$ and $\varepsilon_\mathrm{K}$ are required to be measured quantitatively. However, $\theta_\mathrm{K}$ and $\varepsilon_\mathrm{K}$ always coexist and simultaneously contribute to the detected MOKE signals. In the previous work, these two parameters have to be measured separately by using and adjusting the angle of a quarter-wave-plate both in the traditional MOKE setup \cite{Qiu:2000} and the WM method \cite{He:21}. It not only decreases the measurement efficiency, but also brings an extra error given by the separate measurement processes. {Although the measurement efficiency can be improved by splitting the reflected light into multiple beams and obtaining the signals of $\theta_\mathrm{K}$ and $\varepsilon_\mathrm{K}$ simultaneously through different post-selection processes \cite{Luolan2019,PhysRevA.108.033724}, these two parameters are still measured by two meter states, which introduces added intricacy and raises the expenses of the optical system.}

In this work, we propose a scheme to measure the complex MOKE using WM. For the pre-selection, the polarized light beam is reflected normally from the magnetic film, in which configuration the induced MOKE is widely applied in high-density storage and measurement of the pure PMOKE \cite{Hajjar:1990,Weller:1994}. Next, a glass prism is used to induce the spin Hall effect of light (SHEL), which serves as the weak coupling. After the post-selection, we introduce two {auxiliary} pointers given by the final meter state to characterize the complex MOKE signal. The simultaneous measurement of $\theta_\mathrm{K}$ and $\varepsilon_\mathrm{K}$ in a single WM process is theoretically and experimentally realized. Compared to the previous methods of separately measuring these two parameters, this scheme greatly improves the efficiency and precision. Further, the complex magneto-optical constant $Q$, a coefficient to characterize the intrinsic magneto-optical property of the material, is also calculated in our experiment as an application of this technique. {Other applications that are significant to the exploration of the complex MOKE, such as the manipulation of SHEL through an applied magnetic field\cite{heyu:2019}, are attainable as well in our scheme.}
\begin{figure}[t]
\centering
\includegraphics[width=1\linewidth]{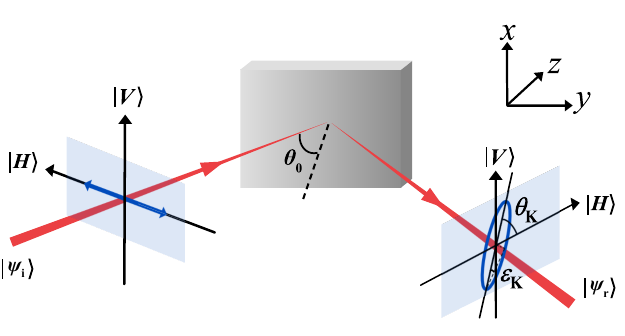}
\caption{Diagram of a {polarized} beam incident on the interface of a magnetic film with the incident angle of $\theta_0$. {Due to MOKE, the polarization of the light undergoes a transformation from the state $|\psi_{\text{i} } \rangle$ to  $|\psi_{\text{r} } \rangle$. $\theta_\mathrm{K}$ and $\varepsilon_\mathrm{K}$ denote the Kerr optical rotation angle and the Kerr ellipticity of the reflected polarization, respectively.}}
\label{MOKE}
\end{figure}

Considering an initialized horizontal polarization $\left|\psi_\text{i}\right\rangle=\left|H\right\rangle$ reaching a magnetic interface from a nonmagnetic medium, as shown in Fig.   \ref{MOKE}, MOKE occurs and the state of the reflected light (the space part is omitted here) gives
\begin{align}
    |\psi^{{1}  }_{\mathrm{r} } \rangle &=r_\mathrm{pp}\left|H\right\rangle+r_\mathrm{sp}\left|V\right\rangle\nonumber\\
&\approx \frac{r_{\mathrm{pp} }}{\sqrt{2} } \left[ \mathrm{e}^{-i\theta_{\mathrm{K} } -\varepsilon_{\mathrm{K} } }|+\rangle +\mathrm{e}^{i\theta_{\mathrm{K} } +\epsilon_{{\mathrm{K} }  } }|-\rangle \right],
\label{R}
\end{align}
where $\left\vert +\right>  =\left( \left\vert H\right>  +i\left\vert V\right>  \right)  /\sqrt{2} $  and   $\left\vert -\right>  =\left( \left\vert H\right>  -i\left\vert V\right>  \right)  /\sqrt{2} $  are the left- and the right-circular polarization states, respectively. In the second line of Eq.   \eqref{R}, the approximation condition $|{r_{\mathrm{sp}}}/{r_{\mathrm{pp}}} |=|\theta_{\mathrm{K} } -i\varepsilon_{\mathrm{K} } |\ll 1$ is used in the calculation. For normal incidence we interested in, the complex MOKE angle can be expressed by \cite{You:1998}
\begin{equation}
\Theta_\mathrm{K}^\mathrm{H}=\theta_\mathrm{K}-i\varepsilon_\mathrm{K}=\frac{r_\mathrm{sp}}{r_\mathrm{pp}}=\frac{i n_0 n_1 Qm_z}{n_1^2 - n_0^2},
\label{Q}
\end{equation}
where $n_0$, $n_1$ are the refractive indices of the nonmagnetic medium, and that of the magnetic medium, respectively. $m_z$ is a direction cosine of the magnetization $\bf M$. $Q=Q_0 \exp (-iq)$ is the complex magneto-optical constant, whose amplitude $Q_0$ is proportional to the magnetization intensity \cite{Yang:1993} and $q$ is its phase factor. {In addition,  it should be emphasized that $r_\mathrm{pp}$ is a constant in this configuration.}

Previous weak measurement scheme for measuring MOKE usually employ the SHEL induced by the magnetic sample as the weak coupling \cite{Qiu:2014,Luo:23,LiTong:23}. However, SHEL only exists with oblique incidence \cite{Luo:2011}, therefore, we consider the case that the light undergoes a second reflection from a prism obliquely with the incident angle of $\theta_\mathrm{i}$ to excite SHEL in the air-glass interface as the weak coupling, which takes the form 
\begin{equation}
\hat{U}_{\text{int}}=\exp(-i\delta \hat{\sigma }_{3} k_{y}),
\end{equation}
where $\hat{\sigma}_3=\left|+\right\rangle \left\langle+\right|-\left|-\right\rangle \left\langle-\right|$, the spin operator of a photon, is the observable of the quantum system being measured, $k_y$ is the transverse component of the wave vector describing the momentum wave function, and $\delta=(1+r_\mathrm{s}/r_\mathrm{p})\cot\theta_\mathrm{i}/k$ represents the photon spin splitting occurring on the air-glass interface\cite{Hosten:2008}, which is used to reflect the strength of the spin-orbit coupling between $\hat{\sigma}_3$ and $k_y$. After that interaction, the polarization and the space parts of the light are entangled, and the total state of the second reflected light gives
\begin{align}
|\psi^{{2}  }_{\mathrm{r} } \rangle  &=\exp(-i\delta \hat{\sigma }_{3} k_{y})\Re_{2} |\psi^{{1}  }_{\mathrm{r} } \rangle\left\vert \phi \right> \nonumber\\
&=\exp(-i\delta \hat{\sigma }_{3} k_{y})|\psi_\mathrm{pre}\rangle\left\vert \phi \right>,
\end{align}
where $\left\vert \phi \right>  =\int dk_{y}  (w/\sqrt{2\pi } )^{1/2}\exp {(-w^{2}k^{2}_{y}/4)}  \left\vert k_y \right>$ is the momentum wave function with Gaussian distribution ($w$ denotes the beam waist), which serves as the meter in the WM scheme. Unlike Eq.   \eqref{eq2} which has non-diagonal elements,
\begin{equation}
\Re_2=\left[ \begin{array}{cc}r_{\mathrm{p} }&0\\ 0&r_{\mathrm{s} }\end{array} \right],  
\end{equation}
is the Jones matrix describing the reflection upon an isotropic non-magnetic dielectric medium {(glass)}, and $|\psi_\mathrm{pre}\rangle=\Re_2|\psi^{{1}  }_{\mathrm{r} } \rangle$ is regarded as the pre-selected state in the WM system. 

Next, the state of the whole systerm undergoes a free evolution $\hat{U}_\mathrm{free}=\exp(-ik_y^2z/2k)$  ($z$ is the free propagation distance)\cite{Hosten:2008}, and during this process, it is post-selected on the state
\begin{equation}
\left|\psi_\mathrm{post}\right\rangle=\left|V\right\rangle=-\frac{i}{{\sqrt{2} }  } (\left| +\right\rangle  -\left| -\right\rangle  ).
\end{equation}
Therefore, the final wave function of the light in the momentum space gives 
\begin{align}
\left|\Phi\right\rangle&=\hat{U}_\mathrm{free}|\psi_\mathrm{post} \rangle\left\langle\psi_\mathrm{post}\right|\exp(-i\delta \hat{\sigma }_{3} k_{y})\left|\psi_\mathrm{pre}\right\rangle\left|\phi\right\rangle \nonumber \\
&=\hat{U}_\mathrm{free}\langle \psi_\mathrm{post} |\psi_\mathrm{pre} \rangle \left[\cos\left( \delta k_{y}\right)  -iA_{w}\sin\left( \delta k_{y}\right)  \right]|\phi\rangle|\psi_\mathrm{post} \rangle ,
\end{align}
\begin{figure*}[t]
\centering
\includegraphics[width=0.99\textwidth]{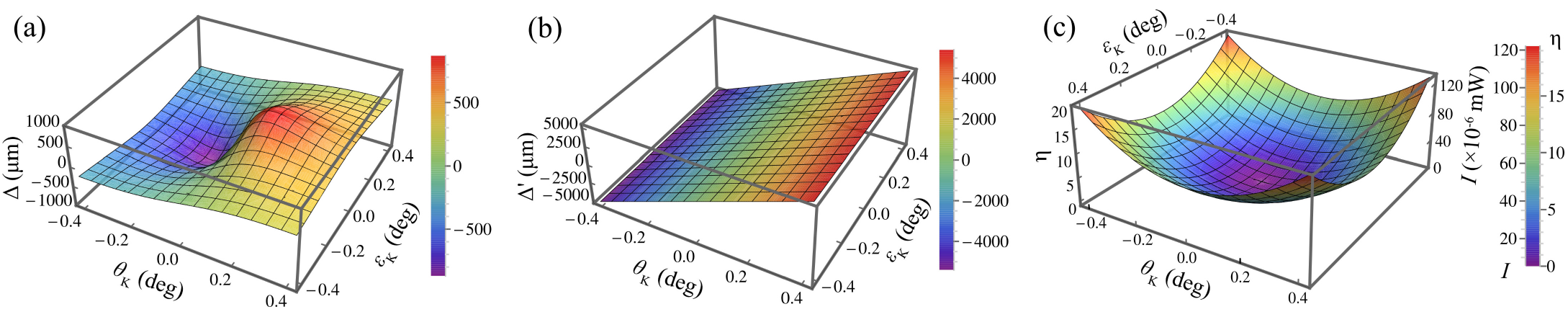}
\caption{{Theoretical dependencies of the pointers on the complex MOKE. (a) and (b) are the amplified shift $\Delta$, and the
modified shift $\Delta '$ with the two MOKE parameters (i.e., $\theta_\mathrm{K}$ and $\varepsilon_\mathrm{K}$). $\Delta '$ shows a higher dynamic measurement range with wider consistent sensitivity compared with $\Delta$ (the standard WM scheme).
(c) shows the curve of the relative intensity change $\eta$ along with the post-selected intensity $I$, where the left and right scales of the color bar represents the values of $I$ and $\eta$, respectively. The parameters taken in the plots correspond to the experiment settings.}}

\label{theory}
\end{figure*}
where
\begin{align}
A_w=\frac{\langle\psi_\mathrm{post}|\hat\sigma_3|\psi_\mathrm{pre}\rangle}{\langle\psi_\mathrm{post}|\psi_\mathrm{pre}\rangle}=\frac{r_\mathrm{p}}{r_\mathrm{s}}\frac{\sinh 2\varepsilon_\mathrm{K}-i\sin 2\theta_\mathrm{K}}{\cos 2\theta_\mathrm{K}-\cosh 2\varepsilon_\mathrm{K}}
\end{align}
is the complex weak value corresponding to the complex MOKE. {What noteworthy is that when only Kerr optical rotation angle $\theta_\mathrm{K}$ (Kerr ellipticity $\epsilon_\mathrm{K}$) appears, the complex $A_w$ degenerates into a pure imaginary (real) one. }

And the amplified shift of the beam, which is the most currently used pointer in WM, can be calculated as

\begin{widetext}
\begin{align}\label{Shift}
\Delta|_{\theta_{\mathrm{k}},\epsilon_{\mathrm{k}} } &=\frac{\langle\Phi|\hat{y}|\Phi\rangle}{\langle\Phi|\Phi\rangle}\nonumber=\frac{\int^{+\infty }_{-\infty } \Phi^{\ast } \left( k_{y}\right)  i\partial {k_{y}} \Phi \left( k_{y}\right)  \mathrm{d}k_{y}}{\int^{+\infty }_{-\infty } \Phi^{\ast } \left( k_{y}\right)  \Phi \left( k_{y}\right)  \mathrm{d}k_{y}} \nonumber\\
&\approx\frac{z}{R_{0}} \frac{2r_{\mathrm{s} }r_{\mathrm{p} }\delta \sin {2\theta_{\mathrm{K} } }  }{\left[r^{2}_\mathrm{s}-r^{2}_\mathrm{p}+\mathrm{e}^{k_{0}\delta^{2} /R_{0}} (r^{2}_\mathrm{s}+r^{2}_\mathrm{p})\right]\cosh {2\varepsilon_{\mathrm{K} }} -    \left[ r^{2}_\mathrm{s}+r^{2}_\mathrm{p}+\mathrm{e}^{k_{0}\delta^{2} /R_{0}} (r^{2}_\mathrm{s}-r^{2}_\mathrm{p})\right]\cos {2\theta_{\mathrm{K} } }  }, 
\end{align}
 \end{widetext}
where $R_0=kw^2/2$ is the Rayleigh distance. The theoretical relationship between $\Delta$ and the complex MOKE is shown in Fig.   \ref{theory}(a). It can be seen that the role of $\varepsilon_\mathrm{K}$ is almost ignored when the complex MOKE signal is weak enough. Thus, the pointer $\Delta$ is used to measure weak $\theta_\mathrm{K}$ in previous literature\cite{He:21}.  {However, the linearity of $\Delta$ with $\theta_\mathrm{K}$ decreases with the increasing of $\theta_\mathrm{K}$, and is affected by the value of $\varepsilon_\mathrm{K}$. For the general cases, the max value of $\theta_\mathrm{K}$ that keeps a good linearity with $\Delta$ are on the order of $10^{-2}$ deg (see the supplementary material for details). This limitation restricts practical applications.} Moreover, it is hard to provide the information about $\varepsilon_\mathrm{K}$ when employing $\Delta$ as the only pointer.

To obtain more information of the complex MOKE, we introduce the post-selected intensity as {an extra pointer:} 
 \begin{widetext}
\begin{align}\label{Intensity}
I |_{\theta_{\mathrm{k}},\epsilon_{\mathrm{k}} }&\propto\langle\Phi|\Phi\rangle = \int^{+\infty }_{-\infty } \Phi^{\ast } \left( k_{y}\right)  \Phi \left( k_{y}\right)  \mathrm{d}k_{y}\nonumber \\ &\propto\left[r^{2}_\mathrm{s}-r^{2}_\mathrm{p}+\mathrm{e}^{k_{0}\delta^{2} /R_{0}} (r^{2}_\mathrm{s}+r^{2}_\mathrm{p})\right]\cosh {2\varepsilon_{\mathrm{K} }} -    \left[ r^{2}_\mathrm{s}+r^{2}_\mathrm{p}+\mathrm{e}^{k_{0}\delta^{2} /R_{0}} (r^{2}_\mathrm{s}-r^{2}_\mathrm{p})\right]\cos {2\theta_{\mathrm{K} } }  .
\end{align}
 \end{widetext}
It contains the information of both $\theta_\mathrm{K}$ and $\varepsilon_\mathrm{K}$ but dependent non-linearly with neither of them. 
 
By combining  Eq.   \eqref{Shift} and \eqref{Intensity}, we can derive {two auxiliary  pointers}, the modified shift $\Delta '$ and the relative intensity change $\eta$, to characterize the complex MOKE. They take the following forms:
\begin{equation}
\Delta '|_{\theta_{\mathrm{k}}}=\frac{I |_{\theta_{\mathrm{k}},\epsilon_{\mathrm{k}} }}{I |_{\theta_{\mathrm{k}}=0,\epsilon_{\mathrm{k}}=0 }}\cdot \Delta \approx \frac{2 z r_\mathrm{s}}{k\delta r_\mathrm{p}}\theta_\mathrm{K},
\label{delta'}
\end{equation} 
 
\begin{equation}
\eta|_{\theta_{\mathrm{k}},\epsilon_{\mathrm{k}} }=\frac{I |_{\theta_{\mathrm{k}},\epsilon_{\mathrm{k}} }-I |_{\theta_{\mathrm{k}}=0,\epsilon_{\mathrm{k}}=0 }}{I |_{\theta_{\mathrm{k}}=0,\epsilon_{\mathrm{k}}=0 }} \approx \frac{r_\mathrm{s}^2 w^2}{r_\mathrm{p}^2 \delta^2}(\theta_\mathrm{K}^2 +\varepsilon_\mathrm{K}^2),
 \label{eta}
 \end{equation}
where $I |_{\theta_{\mathrm{k}}=0,\epsilon_{\mathrm{k}}=0 }$ is the post-selected intensity corresponding to no polarization change. The approximations in Eq.   \eqref{delta'} and \eqref{eta} are valid when $|\theta_{\mathrm{K} } -i\varepsilon_{\mathrm{K} } |\ll 1$ (subtle  MOKE signal), and $|\delta/w|\ll1$ (weak coupling condition). The theoretical dependencies of the {two auxiliary pointers} and the complex MOKE are shown in Fig.   \ref{theory}(b) and Fig.   \ref{theory}(c). It can be seen that the value of $\Delta '$ is {merely} dependent on $\theta_\mathrm{K}$ and the linearity maintains in the whole area. Compared to the $\Delta$ pointer for measuring $\theta_\mathrm{K}$, the $\Delta '$ pointer greatly enlarge the linear response region. $\eta$ is linear to $I$ and determined by both of $\theta_\mathrm{K}$ and $\varepsilon_\mathrm{K}$, which is used for the calculation of $\varepsilon_\mathrm{K}$. Thus, $\theta_\mathrm{K}$ and $\varepsilon_\mathrm{K}$ can be obtained using these two {auxiliary} pointers in a single WM process. It is worthy noting that the values of all the pointers are at the minimum when MOKE is absent. 
 
\begin{figure}[tb]
\centering
\includegraphics[width=0.48\textwidth]{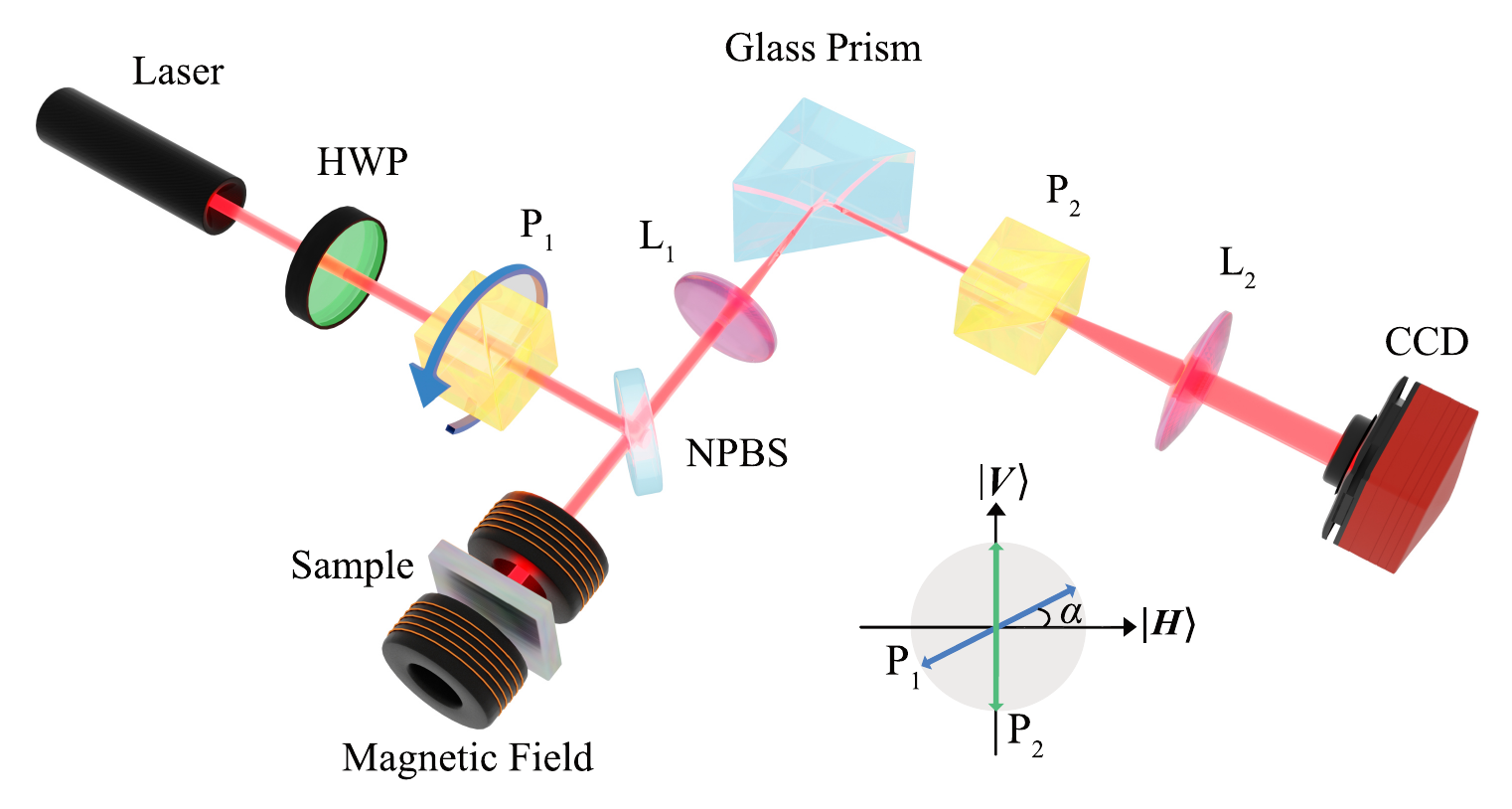}
\caption{Experiment setup for the measurement of the complex MOKE. The light source is a He-Ne laser (wavelength = 623.8 nm), HWP is a half-wave plate, {P$_1$} and {P$_2$} are Glan laser polarizers, NPBS is a non-polarizing beam splitter, L$_1$ and L$_2$ are the lenses with the focal lengths of 50 mm and 250 mm respectively, CCD denotes the charge-coupled device. {The inset shows the polarization states prepared by P$_1$ and P$_2$. The optical axis of P$_1$ deviates an angle of $\alpha$ from $\left|H\right\rangle$, while the optical axis of P$_2$ is parallel to $\left|V\right\rangle$.}}
\label{setup}
\end{figure}

The experimental measurement system is shown in Fig.   \ref{setup}. The light source is a He-Ne laser {(Thorlabs HNL210)} with a wavelength of 632.8 nm, which passes through a half-wave plate (HWP) to adjust the intensity. Next, the light beam passes through a {Glan polarizer (P$_1$, extinction ratio 100 000 : 1)}. Since {standard WM scheme} provides the best sensitivity for the measurement of $\theta_\mathrm{K}$ when the pre-selected state is parallel to $\left|H\right\rangle$, where $\Delta$ is 0. Thus, we adjust P$_1$ and obtain the initial polarization state $\left|\psi_\mathrm{i}\right\rangle=\left|H+\alpha\right\rangle$, where $\alpha=\theta_\mathrm{K0}$ is an auxiliary compensation angle used to offset the initial Kerr rotation angle $\theta_\mathrm{K0}$ (see the supplementary material for details). Then, the beam passes through the non-polarizing beam splitter (NPBS) and normally reaches to the surface of the magnetic thin film, to which the external magnetic field $B$ is applied perpendicular. Then, the reflected light is focused by a lens L$_1$ ($f=50$ mm) and then reflected obliquely from the BK-7 glass prism. The incident angle $\theta_\mathrm{i}=40^{\circ}$ is kept throughout the experiment. With this photon wavelength and incident angle, the refractive index of the glass prism is 1.51. $r_\mathrm{p}$ and $r_\mathrm{s}$ are 0.122 and $-0.282$, respectively. A lens L$_2$ ($f=250$ mm) is used to generate the {free} propagation amplification\cite{Hosten:2008} on the beam and another {Glan polarizer (P$_2$, extinction ratio 100 000 : 1)} is used to prepare $\left|\psi_\mathrm{post}\right\rangle=\left|V\right\rangle$ as the post-selected state. The detection of the reflected beam is finally carried out using a charge coupled device (CCD, {WORK POWER WP-UT400}).
 
\begin{figure}[tb]
\centering
\includegraphics[width=1\linewidth]{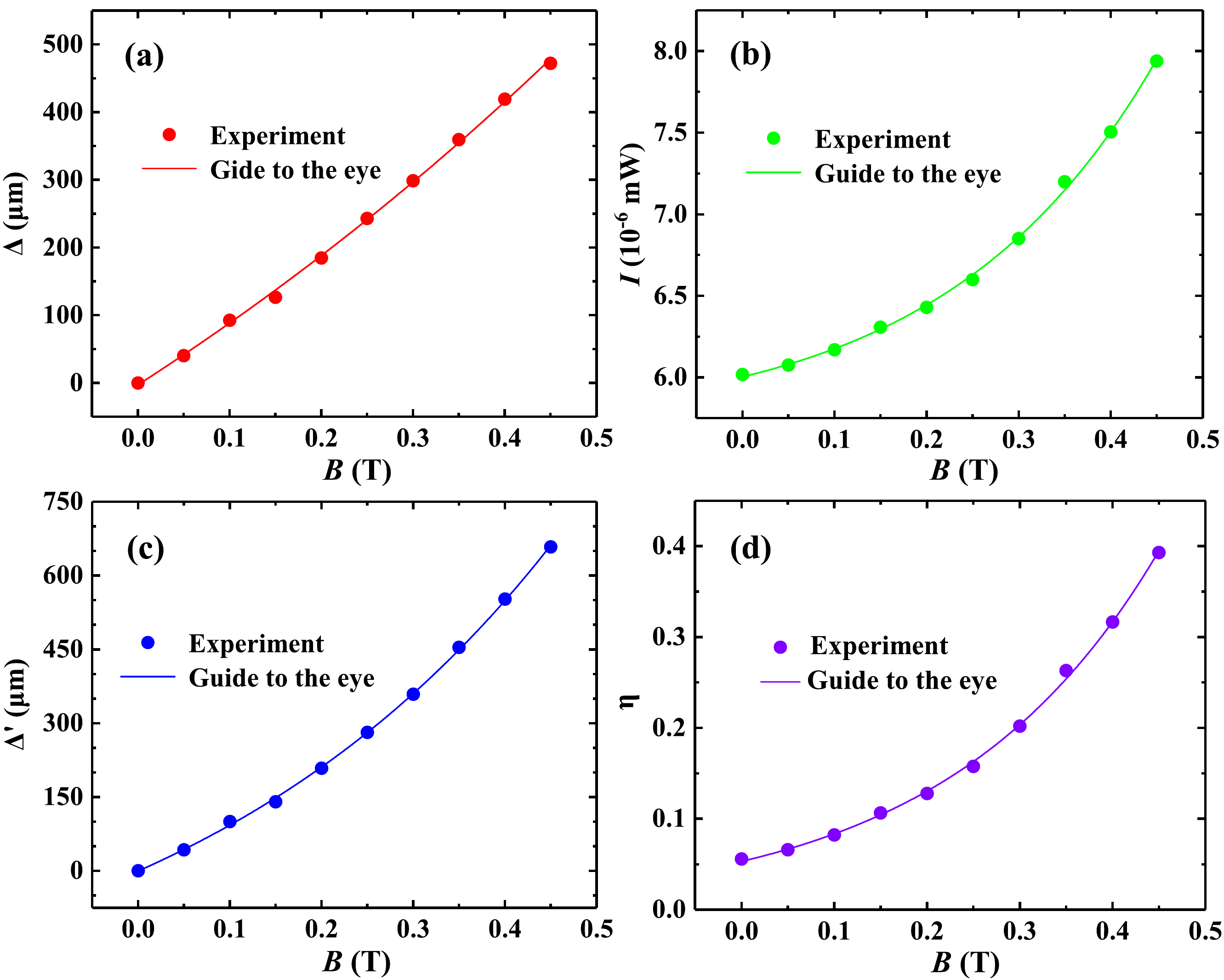}\caption{The dependence of the pointers and the intensity of the magnetic field. (a) and (b) are the experiment results for the measurement of $\Delta$ and $I$, under room temperature. (c) and (d) show the {two auxiliary  pointers} $\Delta '$ and $\eta$ calculated by the measurement results. The scatters are the average values of the measurement, and the curves are the guides to the eye.}
\label{experiment}
\end{figure}

A Ni-Fe film with the thickness of 30 nm that depositing on the silicon substrate using the vacuum thermal evaporation technique
is prepared for the experiment. During the measurement, the applied magnetic field $B$ is changed from 0 to 0.45 T with steps of 0.05 T. At each value of $B$, the amplified shift $\Delta$ and the post-selected intensity $I$ are recorded 30 times with intervals of 1 second by the CCD. The measurement results of $\Delta$ and $I$ are shown in Fig.  \ref{experiment}(a) and \ref{experiment}(b). It can be seen that both $\Delta$ and $I$ increase with the increase of $B$, due to the increase of $\theta_\mathrm{K}$ and $\varepsilon_\mathrm{K}$, which is consistent with our theory. Fig.  \ref{experiment}(c) and \ref{experiment}(d) show the {two auxiliary  pointers} calculated by the experiment results. Here, the post-selected intensity without MOKE $I |_{\theta_{\mathrm{k}}=0,\epsilon_{\mathrm{k}}=0}=5.7\times10^{-6}$ mW is determined by measuring the minimum value of $I$ while changing $B$, under the condition of $\left|\psi_\text{i}\right\rangle=\left|H\right\rangle$. Except for this used method, it can also be determined by employing a quarter-wave-plate to compensate the ellipticity, while adjusting the angle of P$_1$ to compensate the rotation angle.

\begin{figure}[tb]
\centering
\includegraphics[width=0.87\linewidth]{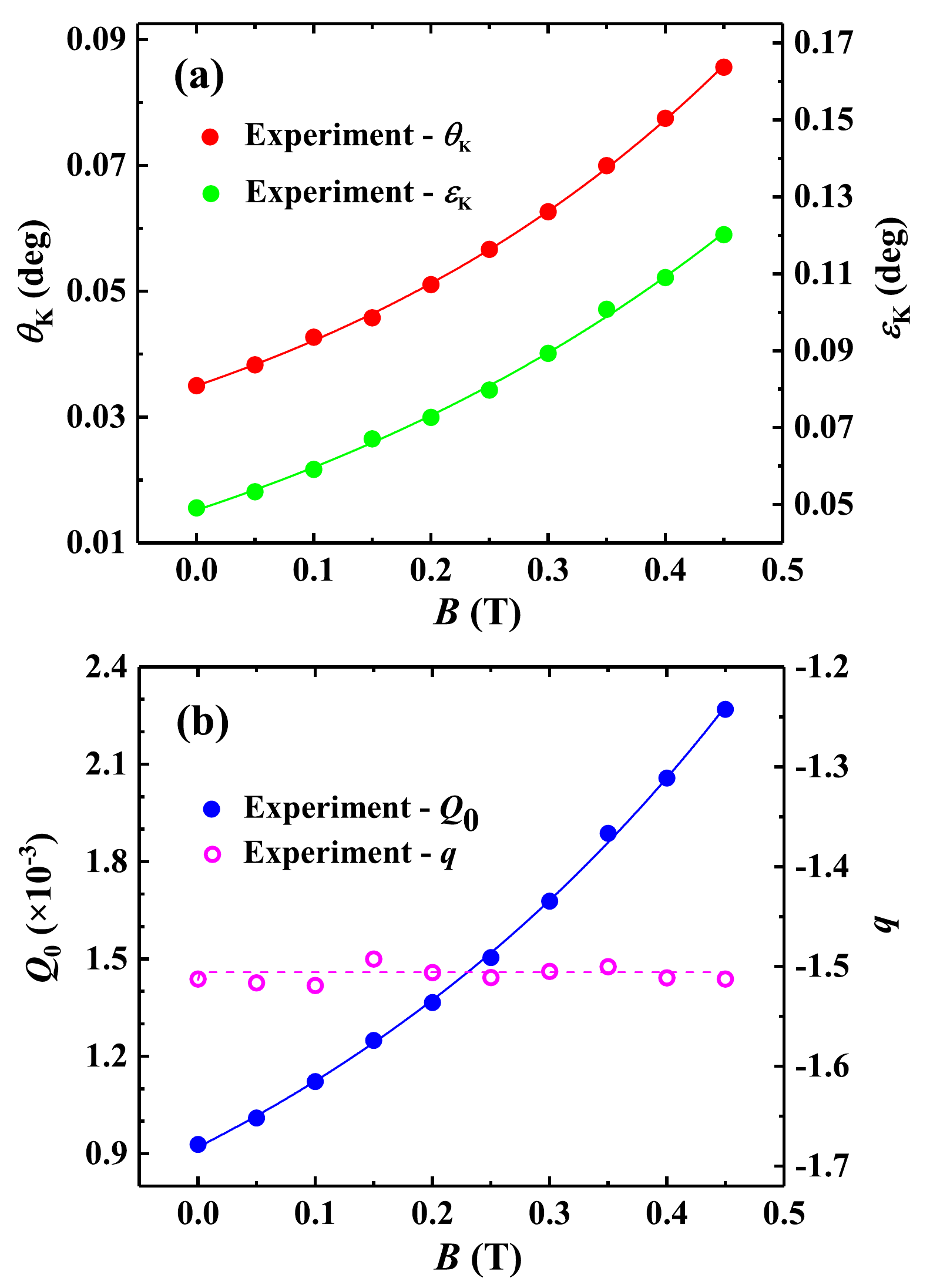}
\caption{The dependence of the complex MOKE parameters {on} the intensity of the magnetic field. (a) $\theta_\mathrm{K}$ and $\varepsilon_\mathrm{K}$, (b) $Q_0$ and $q$. The scatters are the calculated results according to $\Delta'$ and $\eta$, and the curves are the guides to the eye.}
\label{parameters}
\end{figure}

Furthermore, we can characterize the complex MOKE using the {two auxiliary pointers}. According to Eq.  (\ref{delta'}) and Eq.  (\ref{eta}), the expressions of $\theta_\mathrm{K}$ {(with the initial  $\theta_\mathrm{K0}$)} and $\varepsilon_\mathrm{K}$ are
\begin{equation}
\theta_{\mathrm{K} } |_{\Delta^{\prime },\alpha }=\frac{k\delta r_\mathrm{p}}{2 z r_\mathrm{s}}\Delta '+ \theta_\mathrm{K0}|_{\alpha},    
\end{equation}
\begin{equation}
\varepsilon_\mathrm{K}|_{\Delta^{\prime },\eta }=\sqrt{\frac{r_\mathrm{p}^2\delta^2}{r_\mathrm{s}^2 w^2}\eta-(\frac{k\delta r_\mathrm{p}}{2 z r_\mathrm{s}}\Delta ')^2}.
\end{equation}
In this experiment,  $\theta_\mathrm{K0}=\alpha=0.035$ deg. Thus, we can calculate the dependence of the complex MOKE on $B$, which are shown in Fig.  \ref{parameters}(a). It can be seen that both of the two parameters increase with the increasing of $B$. 
The measurement precision of these two parameters can be evaluated by 
\begin{equation}
S_{\theta _{\mathrm{K} }} =\sqrt{(\frac{\partial \theta_\mathrm{K}}{\partial \Delta}S_{\Delta})^2+(\frac{\partial \theta_\mathrm{K}}{\partial I}S_{I})^2+(\frac{\partial \theta_\mathrm{K}}{\partial I_0}S_{I_0})^2},
\end{equation}
\begin{equation}
S_{\varepsilon _{\mathrm{K} }}=\sqrt{(\frac{\partial \varepsilon_\mathrm{K}}{\partial \Delta}S_{\Delta})^2+(\frac{\partial \varepsilon_\mathrm{K}}{\partial I}S_{I})^2+(\frac{\partial \varepsilon_\mathrm{K}}{\partial I_0}S_{I_0})^2},
\end{equation}
where $S_{\Delta}$ $S_{I}$, $S_{I_0}$ are the standard deviations of $\Delta$, $I$, and $I_0$, respectively. In this experiment, $S_{\Delta}\approx2.5$ $\mu$m  and $S_{I_0}\approx S_{I}\approx0.013\times 10^{-6}$ mW are calculated by the 30 times measurement. {Thus, the corresponding standard derivations of $\theta_\mathrm{K}$ and $\varepsilon_\mathrm{K}$, $S_{\theta_\mathrm{K}}\approx S_{\varepsilon_\mathrm{K}}\approx 2.5\times10^{-4}$ deg, are obtained (see the supplementary material for details).} 

Furthermore, we calculate the complex magneto-optical constant $Q$ using this method. From Eq.  (\ref{Q}), $Q$ can be calculated as
\begin{equation}
Q=\frac{(n_1^2-n_0^2)(\varepsilon_\mathrm{K}-i\theta_\mathrm{K})}{n_0n_1m_z}.
\end{equation}
Since the plane of the Ni-Fe film is isotropic, and the applied magnetic field is normal to the film surface, we assume $m_z=1$ during the whole measurement process. The refractive indices of $n_0$ and $n_1$ are 1 and $1.26-0.44i$, respectively, where $n_1$ is measured by the ellipsometer. We depict the dependence of $Q$ and $B$, as shown in Fig.  \ref{parameters}(b). The amplitude $Q_0$ intuitively shows the magnetization performance perpendicular to the film surface, while the phase $q$ remains a constant.

In summary, we propose a method to measure the complex MOKE using WM. By introducing {two auxiliary pointers}, $\Delta'$ and $\eta$, we realize the simultaneous measurement of $\theta_\mathrm{K}$ and $\varepsilon_\mathrm{K}$ with the precision of {$2.5\times10^{-4}$ deg}, which addressed the problem that these two MOKE parameters have to be measured separately using previous methods. Since the information of the two pointers are given by the same meter state, this scheme greatly improve the efficiency and precision for the measurement of the complex optical parameters. {It is worthy noting that a higher measurement precision could be further achieved by combing with the quantum light source or the noise reduction techniques such as the lock-in amplifier.} What's more, since the complex MOKE cause a rotation angle and an ellipticity of the polarization light before it reached to the surface where SHEL occurs, it can be used as a flexible approach to manipulate the spin splitting of SHEL with quantitative analysis. 
\section*{Supplementary Material}
See the supplementary material for (S1) feasibility of introducing the compensation angle $\alpha$, (S2) calculation of the measurement precision of $\theta_\mathrm{K}$ and $\varepsilon_\mathrm{K}$, (S3) the linear response region of the amplified shift $\Delta$ with $\theta_\mathrm{K}$, and (S4) a comparative experimental case to measure the complex MOKE using the standard polarimetry scheme.

\begin{acknowledgments}
This work is supported by National Natural Science Foundation of China (61871451), Sichuan Science and Technology Program (2022YFG0166), and Joint Fund of Bijie City and Guizhou University of Engineering Science (Bijie Science and Technology Union Contract [2023] No.16).
\end{acknowledgments}

\section*{Author Declarations}
\subsection*{Conflict of interest}
The authors have no conflicts to disclose.

\subsection*{Data Availability}
The data that support the findings of this study are available from the corresponding author upon reasonable request.

\nocite{*}
\bibliography{ref}

\end{document}